\documentclass[
aps,
nofootinbib,
prd,
superscriptaddress,
tightenlines,
notitlepage,
twocolumn,
showpacs,
floatfix
]{revtex4-2}
\usepackage{amsmath}
\usepackage{latexsym}
\usepackage{amsfonts}
\usepackage{graphicx}
\usepackage{mathrsfs}
\usepackage{epstopdf}
\usepackage{subfigure}
\usepackage[utf8]{inputenc}
\usepackage{CJK}
\usepackage{float}
\usepackage{color}
\usepackage{hyperref}
\usepackage{diagbox}
\usepackage{xcolor}
\usepackage{bm}
\usepackage{lipsum}
\hypersetup{
    colorlinks=true,
    linkcolor=red,
    citecolor=blue,
}

\def\pt#1{\phantom{#1}}

\newcommand{\bhat}{\hat{\bm{b}}}

\begin{document}

\title{Polarizations of Gravitational Waves in the Bumblebee Gravity Model}

\author{Dicong Liang}
\email[Corresponding author: ]{dcliang@pku.edu.cn}
\affiliation{Kavli Institute for Astronomy and Astrophysics, Peking University, Beijing
100871, China}

\author{Rui Xu}
\affiliation{Kavli Institute for Astronomy and Astrophysics, Peking University, Beijing
100871, China}

\author{Xuchen Lu}
\affiliation{School of Physics, Huazhong University of Science 
and Technology,
Wuhan, Hubei 430074, China}

\author{Lijing Shao}
\email[Corresponding author: ]{lshao@pku.edu.cn}
\affiliation{Kavli Institute for Astronomy and Astrophysics, Peking University, Beijing 100871, China}
\affiliation{National Astronomical Observatories, Chinese Academy of Sciences, Beijing 100012, China}

\begin{abstract}
Lorentz violation modifies the dispersion relation of gravitational waves (GWs), and induces birefringence and anisotropy in propagation. 
Our study shows that Lorentz violation can also activate multiple polarizations of GWs. 
We use the gauge invariants to investigate the polarizations of GWs in the bumblebee gravity model, and obtain the following results.
(i) For a vector background $b^\mu$ with only a nonzero temporal component $b^t$, there are five independent propagating degrees of freedom (DOFs), which is simlar to the Einstein-\ae{}ther theory. 
(ii) The presence of a spatial component in the background defines a preferred spatial direction which breaks rotational symmetry.
We denote $\bhat$ as the direction of the spatial part of the background and $b_s$ as its length.
If GWs propagate along $\bhat$, the polarization content is similar to the purely timelike case.
(iii) If the propagation direction of GWs is separated by an angle $\beta$ to $\bhat$, and $\beta=\arccos(b^t/b_s)$, there are only two tensor polarizations.
(iv) If $\beta\neq \arccos(b^t/b_s)$, there are only two independent DOFs, and the vector  and  scalar modes degenerate with the tensor modes. 
The tensor perturbations can activate a mixture of all six polarizations simultaneously.
Finally, we point out the difference in GWs between the bumblebee gravity model and the minimal Standard-Model Extension framework in the linearized regime.
Current observations have placed stringent constraints on the anisotropy induced by the background, while our theoretical study still reveals some novel phenomena and provides more understanding about the interaction between the Lorentz-violating vector field and gravity.
\end{abstract}
\maketitle

\allowdisplaybreaks 

\section{Introduction}

With the increasing number of confident gravitational-wave (GW) detections by the LIGO Scientific Collaboration, the Virgo Collaboration, and the KAGRA Collaboration \cite{LIGOScientific:2016aoc,LIGOScientific:2018mvr,LIGOScientific:2020ibl,LIGOScientific:2021usb,LIGOScientific:2021djp}, it raises more and more interests in the GW community to test general relativity (GR) with GWs. 
One of the most important properties of GWs is the polarization.
There are only two polarizations in GR, that are the plus mode and the cross mode.
But in general metric gravity theories, there can be up to six polarizations \cite{Eardley:1973zuo,Eardley:1973br}. 
For example, in the Brans-Dicke theory, the massless scalar produces an extra breathing mode \cite{Eardley:1973zuo}. 
In $f(R)$ theory and Horndeski theory, the massive scalar field activates a mixture of the breathing mode and the longitudinal mode \cite{Liang:2017ahj,Moretti:2019yhs,Hou:2017bqj}. 
The vector-x mode and the vector-y mode are produced by the
dynamic vector field in the Einstein-\ae{}ther theory or the tensor-vector-scalar (TeVeS) theory \cite{Jacobson:2004ts,Sagi:2010ei,Gong:2018cgj}.  
The polarizations of GWs in other modified gravity theories were further analyzed in Refs.~\cite{Gong:2018vbo,Wagle:2019mdq,Liu:2019cxm,Bombacigno:2019did,Lu:2020eux,Dong:2021jtd,Farrugia:2018gyz,Soudi:2018dhv,Capozziello:2019msc,Capozziello:2020vil,Bahamonde:2021dqn,Capozziello:2021bki,Tachinami:2021jnf}.

The first polarization analysis with the real observational data was performed for GW150914 \cite{LIGOScientific:2016lio}. Due to the co-alignment and the limited number of the detectors, this coherent Bayesian analysis is very inclusive. 
Later, similar analyses were performed for the three-detector observations GW170814 \cite{LIGOScientific:2017ycc}, GW170817 \cite{LIGOScientific:2018dkp,Takeda:2020tjj}, and other events in the first GW transient catalog (GWTC-1) \cite{LIGOScientific:2019fpa}.
In addition to the constraints on pure polarization in the previous literature, an extended study on constraints on a mixture of tensor and scalar polarizations was performed in Ref. \cite{Takeda:2021hgo}. 
Null stream \cite{Guersel:1989th,Chatterji:2006nh} can also be used to test the existence of extra polarizations \cite{Chatziioannou:2012rf,Hagihara:2018azu,Hagihara:2019rny,Wong:2021cmp}, 
which has been applied to the event GW170817 \cite{Hagihara:2019ihn,Pang:2020pfz} and events in GWTC-2 and GWTC-3 \cite{LIGOScientific:2020tif,LIGOScientific:2021sio}.
The null stream method can also be adopted by the space-based detectors in the future \cite{Zhang:2021fha, LISA:2022kgy}.

In this work, we analyze the polarizations of GWs in a Lorentz-violating gravity
theory called the bumblebee gravity model, which was considered as a specific
model in the Standard-Model Extension (SME) framework
\cite{Kostelecky:1989jw,Kostelecky:1989jp,Kostelecky:2003fs}.  The SME is a
general framework which contains the Lagrange densities for both the Standard
Model and GR, incorporating additional background fields describing arbitrary
Lorentz violation and CPT violation
\cite{Colladay:1996iz,Colladay:1998fq,Bailey:2006fd}.  For the theoretical
studies on SME, readers are referred to Refs.
\cite{Bailey:2004na,Mattingly:2005re,Kostelecky:2007zz,Kostelecky:2008in,Kostelecky:2008ts,Kostelecky:2009zp,Kostelecky:2010ze,Kostelecky:2011gq,Kostelecky:2013rta}
and references therein.  Within the framework of SME, different types of
terrestrial experiments and astrophysical observations placed constraints on the
Lorentz-violating parameters
\cite{Wolf:2006uu,BaBar:2007run,Hohensee:2009zk,Bocquet:2010ke,D0:2012rbu,Hohensee:2013cya,MINOS:2008fnv,MINOS:2010kat,IceCube:2017qyp,Hohensee:2011wt,Flowers:2016ctv,Muller:2007zz,Parker:2011eb,Battat:2007uh,Bourgoin:2016ynf,Bourgoin:2017fpo,Kostelecky:2006ta,Kostelecky:2013rv,Kostelecky:2008be,Shao:2014oha,
Shao:2022aax,Kostelecky:2016kfm}.

As a specific model, the background field inducing Lorentz violation and CPT violation in the bumblebee model is a single vector field called the bumblebee field.
The presence of a vector background implies a preferred direction. Then physical Lorentz-symmetry breaking occurs if particles or fields have observable interactions with the background \cite{Kostelecky:2003fs}.
The bumblebee model offers rich physical insights. It can reduce to the Einstein-Maxwell theory, the Nambu model \cite{Nambu:1968qk},
or the Will-Nordtvedt theory \cite{Will:1972zz,Hellings:1973zz} in some special limits.
The massless Nambu-Goldstone modes and massive Higgs-type modes were investigated in Refs.~\cite{Bluhm:2004ep,Bluhm:2007bd}.
Some vacuum solutions of the bumblebee model were studied in Ref.~\cite{Bertolami:2005bh}, and the parametrized post-Newtonian (PPN) parameters were computed.
Schwarzschild-like, Kerr-like black hole and traversable wormhole solutions were derived in
Refs.~\cite{Casana:2017jkc,Li:2020dln,Ovgun:2018xys}, and more general static black hole solutions and their properties are discussed by \citet{Xu:2022frb}.
\citet{Bluhm:2008yt} performed the Hamiltonian constraint analysis for the bumblebee model in flat spacetime to investigate the physical degrees of freedom (DOFs) of the vector field. 
Hamiltonian analyses were performed on modified gravity theories with nondynamical background fields in SME framework in Refs. \cite{Reyes:2021cpx,Reyes:2022mvm}.
In this paper, we will use gauge invariants to analyze the DOFs for both the vector field and the metric field in the bumblebee model.

The first study of plane wave solutions in Lorentz-violating gravity was performed for Chern-Simons gravity, which shows that the two tensor polarizations carry different intensities while their speed remains unmodified \cite{Jackiw:2003pm}.
The modified dispersion relations for GWs were considered in several models in Ref.~\cite{Ferrari:2006gs}, where the Lorentz-violating term was introduced in the linearized gravity.
More general linearized-gravity extension with Lorentz and diffeomorphism violations was constructed in Ref.~\cite{Kostelecky:2016kfm}, and then was developed in Refs.~\cite{Kostelecky:2017zob,Mewes:2019dhj,Nascimento:2021rlg}. 
For general Lorentz-violating theories, modifications on GWs include dispersion, birefringence, and anisotropy.
Based on these effects, various authors have performed analyses on the GW data to constrain Lorentz violation \cite{Wang:2021ctl,ONeal-Ault:2021uwu,Shao:2020shv,Wang:2021gqm,Zhao:2022pun,Niu:2022yhr,Haegel:2022ymk}.
In Refs.~\cite{Maluf:2013nva,Amarilo:2018zqg}, the modified graviton propagator in the bumblebee gravity model was derived and the dispersion relation of GWs was obtained. But the kinetic term and the potential of the bumblebee field were not included in the calculation. 
In addition, these papers considered modifications on the tensorial GWs but have not considered modifications on the polarization content.
The Einstein-\ae{}ther theory and the TeVeS theory are Lorentz-violating theories, with the Lagrangian containing the most general kinetic term quadratic in derivatives \cite{Jacobson:2000xp,Bekenstein:2004ne}. 
The polarizations in these two theories were studied by \citet{Gong:2018cgj}, where the background vector field is set to be purely timelike, such that the tensor, vector, and scalar perturbations decouple from each other. 

In this paper, we will discuss the linearized field equations for both timelike and spacelike background vector fields in the bumblebee gravity model.
With the linearized field equations, we can obtain both the dispersion relation and the polarization content simultaneously in this Lorentz-violating theory. 
To our knowledge, we are the first to investigate the polarizations of GWs with a nonzero spatial component of the background vector field, 
which defines a special direction and causes anisotropy.
For a purely timelike bumblebee background $b^\mu=(b^t,0,0,0)$, we find that the polarization content is similar to that in the Einstein-\ae{}ther theory.
But when the spatial component of the background field is included, the rotation symmetry is broken and the polarization content can be different for GWs coming from different directions.
We denote $\bhat$ as the direction of the spatial part of the vector background and $b_s$ as its length.
For GWs propagating along $\bhat$, there are five DOFs, which is similar to the purely timelike case.
If the angle between the propagation direction of  GWs and $\bhat$ is $\beta$, then $\beta^\star\equiv \arccos(b^t/b_s)$
introduces a special spatial direction. There are only the plus mode and the cross mode. For $\beta \notin \{ 0, \pi, \beta^\star \}$, there are only two DOFs in GWs, with the vector and scalar perturbations mixing into the tensor perturbations, i.e.\ they are not independent modes. As far as we know, the phenomenon that a mixture of all the six basic polarizations can be activated by the tensor perturbations is new.

We also compare the polarizations of GWs in the bumblebee model to those in the linearized regime of the SME framework presented by \citet{Bailey:2006fd}.
Following them, we can get an effective bumblebee-like SME model and then apply the same polarization analysis.
In this model, there are always two independent propagating DOFs, regardless of the direction of GWs, which is different from the original bumblebee model. It provides new insights into the relation between the SME framework and the bumblebee gravity model.

The paper is organized as follows.
We give a brief introduction to the bumblebee model and write the linearized field equations in terms of gauge invariants in Sec.~\ref{bumblelinear}.
Polarization analysis is applied to the bumblebee model for various vector backgrounds in Sec.~\ref{bumblepolar}. 
Then in Sec.~\ref{comparison}, we introduce the bumblebee-like SME model and compare its polarization content with the bumblebee gravity model.
Section~\ref{discussion} is for discussions and conclusions.
We put details of our calculation in Appendices
\ref{gi}, \ref{bbbappendix}, and \ref{bsmeappendix}, for the fluency of the paper. Throughout the paper, we adopt the speed of light $c=1$.

\section{Linearized Regime}
\label{bumblelinear}

The bumblebee gravity model introduces a dynamical vector field coupled to gravity, and Lorentz violation arises from the nonzero vacuum expectation value of the vector field. 
The action is given by \cite{Kostelecky:2003fs}:
\begin{equation}
\label{bumblebeeaction}
\begin{aligned}
    S=\int \sqrt{-g} d^4x \big[& \frac{1}{2\kappa} (R+\xi B^\mu B^\nu R_{\mu\nu})-\frac{1}{4}B_{\mu\nu}B^{\mu\nu} \\
   & -V(B^{\mu} B_{\mu} \pm b^2) \big] +S_m ,
\end{aligned}
\end{equation}
where the field strength is
\begin{equation}
    B_{\mu\nu}=D_\mu B_\nu -D_\nu B_\mu ,
\end{equation}
$\kappa=8\pi G$, $\xi$ is a real coupling constant controlling the
curvature-coupling term, and $b^2$ is a real positive constant.
The potential provides a nonzero vacuum expectation value, $b^\mu$, for the bumblebee field. We have $b^\mu b_\mu = \mp b^2$ at the minimum of the potential $V(\cdot)$, where the sign depends on whether the field is timelike or spacelike.
In this paper, we consider a smooth potential
\begin{align}
 V=\lambda (B^\mu B_\mu \pm b^2)^2/2  ,  
\end{align}
where $\lambda$ is a real constant. 
Such a quadratic form is simple and is generally used in literature. It also indicates the existence of a nonvanishing vector background at the minimum of the potential. The analysis in this paper can be applied to other form of potential directly.
The field equation for the vector field in vacuum is given by
\begin{align}
D^\mu B_{\mu\nu} &= 2V' B_\nu -\rho B^\mu R_{\mu\nu},
\label{vectoreq}
\end{align}
where we have defined
$\rho=\xi/\kappa$, and the prime denotes the derivative with respect to the argument. 
The metric field equation in vacuum is:
\begin{align}
G_{\mu\nu} &=\kappa ( T_{\mu\nu}^B +T_{\mu\nu}^\xi) ,
\label{tensoreq}
\end{align}
where
\begin{align}
T^B_{\mu\nu} &= -B_{\mu\alpha}{B^{\alpha}}_{\nu}-\frac{1}{4}g_{\mu\nu}B_{\alpha\beta}B^{\alpha\beta}-g_{\mu\nu}V+2V'B_\mu B_\nu ,\\
T^\xi_{\mu\nu} &= \rho \left[  
        \frac{1}{2}g_{\mu\nu}B^\alpha B^\beta R_{\alpha\beta} -B_\mu B^\alpha R_{\alpha\nu}  -B_\nu B^\alpha R_{\alpha\mu}  \right. \nonumber \\
       & \hspace{0.5cm}  +\frac{1}{2}D_\alpha D_\mu(B^\alpha B_\nu)+\frac{1}{2}D_\alpha D_\nu(B^\alpha B_\mu) \nonumber \\
       & \hspace{0.5cm} \left. -\frac{1}{2}D^\alpha D_\alpha(B_\mu B_\nu)-\frac{1}{2}g_{\mu\nu}D_\alpha D_\beta (B^\alpha B^\beta)   \right] .
\end{align}
We can also obtain the following covariant conservation law
by taking the covariant derivatives of Eq. \eqref{vectoreq} and Eq. \eqref{tensoreq}:
\begin{align}
\label{Vconserv}
2 D^\nu (B_\nu V') &= \rho D^\nu( B^\alpha R_{\alpha\nu} ) , \\
\label{gconserv}
D^\mu T^B_{\mu\nu}& =\rho D^\beta (R_{\alpha\beta} B^\alpha B_\nu ) -\frac{1}{2}\rho R^{\alpha\beta}D_{\nu}(B_\alpha B_\beta) .
\end{align}

To linearize the field equations, 
we separate the metric and the bumblebee field into the background and the perturbation:
\begin{subequations}
\begin{align}
g_{\mu\nu} &= \eta_{\mu\nu}+h_{\mu\nu} , \\
B^\mu &= b^\mu +\tilde{B}^\mu , \\
B_\mu &= b_\mu +\tilde{B}_\mu + b^\alpha  h_{\alpha\mu} ,
\end{align}
\end{subequations}
where we assume that $b^\mu$ is constant, i.e.\ $\partial_\alpha b^\mu =0$, as in Ref.~\cite{Bailey:2006fd}.
Then keeping terms linear in $\tilde{B}_\mu$ and $h_{\mu\nu}$ we have
\begin{align}
\label{Vp}
V'=\lambda(2b^\alpha \tilde{B}_\alpha +b^\alpha b^\beta h_{\alpha\beta} ), 
\end{align}
and the linearized field equations become
\begin{align}
\label{Bumvfieldo}
C_\nu & = 2b_\nu V' -\rho b^\alpha R_{\alpha\nu}    ,\\
\label{Bumgfieldo}
G_{\mu\nu} &= \xi\bigg(W_{\mu\nu} -\frac12 b_\mu C_\nu -\frac12 b_\nu C_\mu +\frac12 \bar{A}_{\mu\nu} -\frac12 \eta_{\mu\nu} \bar{D} \bigg)
\nonumber \\
& \hspace{0.5cm} +2 \kappa b_\mu b_\nu V' ,
\end{align}
where we denote
\begin{subequations}
\begin{align}
C_\nu & \equiv  \Box\tilde{B}_\nu - \tilde{B}^\alpha_{\pt\alpha ,\alpha\nu} +b^\rho (\Box h_{\rho\nu} - {h^\alpha}_{\rho,\alpha\nu} ) \ , 
\\
W_{\mu\nu}  & \equiv \frac12 \eta_{\mu\nu} b^\alpha b^\beta R_{\alpha\beta} -b_\mu b^\alpha R_{\alpha\nu} -b_\nu b^\alpha R_{\alpha\mu} \ ,
\\
\bar{A}_{\mu\nu} & \equiv  b^\alpha \tilde{B}_{\mu,\alpha\nu}   +b^\alpha \tilde{B}_{\nu,\alpha\mu} + b^\alpha b^\rho h_{\mu\nu,\rho\alpha} \ , 
\\
\bar{D} & \equiv 2b^\beta \tilde{B}^\alpha_{\pt\alpha,\alpha\beta} +\frac{1}{2}b^\alpha b^\rho h_{,\alpha\rho}- \frac{1}{2}b^\alpha b^\rho\Box h_{\alpha\rho} +b^\beta b^\rho {h^\alpha}_{\rho,\beta\alpha} \  .
\end{align}
\end{subequations}
Here, the comma denotes the partial derivative.
The linearization of Eq. \eqref{Vconserv} in the momentum space can be simplified to
\begin{equation}
\label{linvconserv}
\begin{aligned}
& p^\nu \left(2\kappa b_\nu V' -\frac{1}{2}\xi b_\nu R +\frac{1}{2}\xi b_\nu R  -\xi b^\alpha  R_{\alpha\nu} \right) \\
=& \frac{1}{2\kappa}b_\nu p^\nu (4 V' -\rho R) =0 ,
\end{aligned}
\end{equation}
which implies
\begin{align}
    b_\nu p^\nu=0,
\label{bp=0}
\end{align}
or
\begin{equation}
    V' =\frac{\rho R}{4} .
\label{vprime}
\end{equation}
Here, $p^\nu$ is the four momentum.
It is consistent with the result of \citet{Bailey:2006fd}, where they obtained the solution of $\tilde{B}^\mu$ in the momentum space first,
\begin{equation}
\begin{aligned}
\tilde{B}^\mu (p) =& -h^{\mu\alpha}b_\alpha
+\frac{p^\mu b^\alpha b^\beta h_{\alpha\beta}}{2b^\alpha p_\alpha}  
-\frac{\rho b^\mu R}{2 p^\nu p_\nu} 
+\frac{\rho p^\mu R}{8\lambda b^\alpha p_\alpha} \\
& +\frac{\rho p^\mu b^\alpha b_\alpha R}{2p^\nu p_\nu b^\alpha p_\alpha}
+\frac{\rho b_\alpha R^{\alpha\mu} }{ p^\nu p_\nu} 
-\frac{\rho p^\mu b^\alpha b^\beta R_{\alpha\beta} }{  p^\nu p_\nu b^\alpha p_\alpha } ,
\end{aligned}
\end{equation}
and then substituted it into Eq.~\eqref{Vp}.
As for the conservation law for the energy-momentum tensor, the linearized Eq.~\eqref{gconserv} is equivalent to Eq.~\eqref{linvconserv} and provides no further simplifications.
Thus, the covariant conservation law in Eqs.~\eqref{Vconserv} and \eqref{gconserv} implies that either GWs propagate perpendicularly to the vector background $b^\mu$, i.e. Eq.~\eqref{bp=0}, or there is a constraint on the relationship between the vector field and the curvature, i.e. Eq.~\eqref{vprime}. 

To write the field equations in terms of gauge invariants, we first follow Refs. \cite{Jackiw:2003pm,Flanagan:2005yc} to decompose the perturbations into irreducible pieces:
\begin{align}
  h_{tt} &=2\phi_h ,
  \nonumber \\
  h_{ti} &=\beta_i +\partial_i \gamma ,
  \nonumber \\
  h_{ij} &=h_{ij}^{\text{TT}}+\frac{1}{3}\delta_{ij}H +\partial_{(i}\epsilon_{j)} +\left( \partial_i\partial_j -\frac{1}{3}\delta_{ij}\nabla^2 \right) \zeta , 
  \nonumber \\
  \tilde{B}^t &= \phi_b , 
  \nonumber \\
\tilde{B}^i &= \mu^i +\partial^i \omega ,
\end{align}
together with the constraints 
\begin{align}
\label{constraints}
   \partial_i \beta_i &=0 ,
   \nonumber \\
   \partial_i \epsilon_i &=0 ,
   \nonumber \\
   \partial_i h_{ij}^{\text{TT}} &=0 ,
   \nonumber \\
   \delta^{ij} h_{ij}^{\text{TT}} &=0 , 
   \nonumber \\
   \partial_i \mu^i  &=0 .
\end{align}
The transformation under infinitesimal particle diffeomorphisms is given by \citet{Bluhm:2007bd}:
\begin{align}
h_{\mu\nu} & \to  h_{\mu\nu}-\partial_\nu \xi_\mu -\partial_\mu \xi_\nu ,
\nonumber \\
\tilde{B}^\mu & \to \tilde{B}^\mu +b^\nu \partial_\nu \xi^\mu  .
\end{align}
It is easy to show that $h_{ij}^{\text{TT}}$ and the following combinations are gauge invariants \cite{Flanagan:2005yc}: 
\begin{align}
\Phi & \equiv -\phi_h +\gamma_{,t} -\frac{1}{2} \zeta_{,tt} \ ,
\nonumber \\
\Theta & \equiv \frac{1}{3} ( H -\nabla^2 \zeta ) \ ,
\nonumber \\
\Xi_i & \equiv \beta_i -\frac{1}{2}\epsilon_{i,t} \  ,
\nonumber \\
\Omega &\equiv \phi_b -b^t\gamma_{,t} -b^i \gamma_{,i} +\frac{1}{2}b^t \zeta_{,tt} +\frac{1}{2}b^i \zeta_{,ti} \ ,
\nonumber \\
\Psi &\equiv \omega +\frac{1}{2} b^t \zeta_{,t}+\frac{1}{2} b^i \zeta_{,i} \ ,
\nonumber \\
\Sigma_i &\equiv \mu_i +\frac{1}{2} b^t \epsilon_{i,t} +\frac{1}{2}b^j \epsilon_{i,j} \  .
\end{align}

In this paper, we focus on the propagation of GWs, thus we only consider the field equations in vacuum. 
Without losing generality, we further adopt a coordinate system in which GWs propagate along the $+z$ axis to simplify the calculation. We call it the GW coordinate system in which we have $h_{\mu\nu}=h_{\mu\nu}(t,z)$ and $\tilde{B}^\mu=\tilde{B}^\mu(t,z)$.
Using the constraints in Eq. \eqref{constraints},
we now have ten nonzero gauge invariants.
They are the four scalar perturbations,
$\Psi(t,z)$, $\Omega(t,z)$, $\Theta(t,z)$, $\Phi(t,z)$,
the four vector perturbations,
$\Sigma_x(t,z)$, $\Sigma_y(t,z)$, $\Xi_x(t,z)$, $\Xi_y(t,z)$,
and the two tensor perturbations,
$h_{xx}^{\text{TT}}=-h_{yy}^{\text{TT}}\equiv h_+(t,z)$, $h_{xy}^{\text{TT}}\equiv h_\times(t,z)$. 
Now, we can express the field equations in terms of the linear combinations of the gauge invariants as shown in Appendix~\ref{gi}.

\section{Polarization Content}
\label{bumblepolar}

Under the assumption that the bumblebee field does not couple to conventional matter, the observable effects of GWs are manifested in the geodesic deviation equation:
\begin{equation}
    \frac{d^2 L^i}{dt^2} =-R_{titj} L^j,
\end{equation}
where the ``electric'' component of the Riemann tensor $R_{titj}$ can be written as 
\begin{equation}
	R_{titj}=-\frac{1}{2}h_{ij,tt}^{\text{TT}}+\Phi_{,ij}+\frac{1}{2}\Xi_{i,tj}+\frac{1}{2}\Xi_{j,ti}-\frac{1}{2}\delta_{ij}\Theta_{,tt} \,.
\end{equation}
Since $\Psi$, $\Omega$, $\Sigma_i$ are not included in $R_{titj}$, they do not affect the geodesic deviation equation. Thus they are not counted as DOFs in GWs.

In the GW coordinate system, the six polarizations are defined by \cite{Eardley:1973zuo}:
\begin{align}
\hat{P}_+ &\equiv -R_{txtx}+R_{tyty} = h_{+,tt} , 
\nonumber\\
\hat{P}_\times &\equiv 2R_{txty} =-h_{\times,tt} ,
\nonumber\\
\hat{P}_x &\equiv R_{txtz} =\frac{1}{2}\Xi_{x,tz} ,
\nonumber\\
\hat{P}_y &\equiv R_{tytz} =\frac{1}{2}\Xi_{y,tz} ,
\nonumber\\
\hat{P}_b &\equiv R_{txtx}+R_{tyty} =-\Theta_{,tt} ,
\nonumber\\
\hat{P}_l &\equiv R_{tztz} =\Phi_{,zz} -\frac{1}{2}\Theta_{,tt} .
\end{align}
It can be shown that  in GR, only the tensor perturbations $h_{+}$ and $h_{\times}$ satisfy wave equations, while the other perturbations are determined by Poisson-type equations \cite{Flanagan:2005yc}.
Thus, only the plus mode and the cross mode propagate in GR. 
Now, we will discuss the polarization content and DOFs of GWs in the presence of the bumblebee background $b^\mu$ for three cases:
\begin{itemize}
	\item Case (I): $b_x^2+b_y^2\neq0$ and $b^t\neq b^z$;
	\item Case (II): $b^x=b^y=0$ and $b^t\neq b^z$;
	\item Case (III): $b^t=b^z$; 
\end{itemize}
These cases are separately listed based on their different properties in GW polarizations.

\subsection{ Case (I): $b_x^2+b_y^2\neq0$ and $b^t\neq b^z$}
\label{bumgeneral}

In this case, the tensor,  vector, and scalar perturbations couple to each other. We introduce higher-order derivatives to separate the perturbations, as shown in Appendix \ref{bbbappendix}.
There are two sets of wave solutions for the field equations.

For the first solution, the wave equations for the tensor modes are given by
\begin{subequations}
\label{bumso1a}
\begin{align}
\label{bumso1p}
-(1-\xi b_t^2)h_{+,tt} +2\xi b^t b^z  h_{+,tz}
+(1+\xi b_z^2) h_{+,zz} &=0 ,  
\\
\label{bumso1c}
-(1-\xi b_t^2)h_{\times,tt} +2\xi b^t b^z  h_{\times,tz}
+(1+\xi b_z^2) h_{\times,zz} &=0 ,
\end{align}
\end{subequations}
and the corresponding dispersion relation is
\begin{equation}
\label{c1tensordp}
\omega =\frac{\sqrt{1-\xi b_t^2 +\xi b_z^2}-\xi b^t b^z}{1-\xi b_t^2} k . 
\end{equation} 
The vector and scalar perturbations are not independent, given in terms of $h_+$ and $h_\times$ by
\begin{widetext}
\begin{subequations}
\label{bumso1b}
\begin{align}
\label{s1th}
\Theta &= -\frac{\xi(b_x^2 -b_y^2) h_+ +2\xi b^x b^y  h_\times  }{2 -\xi (2 b_t^2 -b_x^2 -b_y^2 -2b_z^2) } ,
\\
\label{s1phi}
\Phi &=  \frac{2(1+\xi b_s^2) -4\xi b^t b^z  v_p
+2\xi(2b_t^2 -3b_x^2 -3b_y^2) v_p^2}
{2\xi(b_x^2 +b_y^2) 
[2- \xi (2b_t^2 -b_x^2 -b_y^2 -2b_z^2)]}
 [ \xi (b_x^2-b_y^2)h_+ +2\xi b^x b^y h_\times  ] ,
\\
\label{s1xix}
\Xi_x &= \frac{ (\xi b^t-\xi b^z v_p)
[ 2b^x (1 -\xi (b_t^2 -b_y^2 -b_z^2)) h_+ 
+b^y (2 -\xi(2b_t^2 +b_x^2 -b_y^2 -2b_z^2)) h_\times]}
{(1-\xi b_t^2 +\xi b_z^2)
[ 2 -\xi (2 b_t^2 -b_x^2 -b_y^2 -2b_z^2)] }  ,
\\
\label{s1xiy}
\Xi_y &= \frac{ (\xi b^t-\xi b^z v_p)
[ -2b^y (1 -\xi(b_t^2 -b_x^2 -b_z^2)) h_+ 
+b^x (2 -\xi(2b_t^2 -b_x^2 +b_y^2 -2b_z^2 )) h_\times] }
{(1-\xi b_t^2 +\xi b_z^2)
[ 2 -\xi (2b_t^2 -b_x^2 -b_y^2 -2b_z^2) ] } ,
\end{align}
\end{subequations}
\end{widetext}
where $b_s^2=b_x^2+b_y^2+b_z^2$ and $v_p$ is the phase velocity.
Since we have not detected any Lorentz-violating effects in the gravitational experiments, the coupling terms should be very small, i.e. $ \big| \xi b^\mu b^\nu \big| \ll 1$.
Thus, up to leading order, the extra polarizations are 
\begin{subequations}
\begin{align}
\hat{P}_x &=  \frac12 \xi (b^t - b^z) (b^x h_{+,tz} 
+b^y h_{\times,tz} )  
+\mathcal{O}((\xi b^\mu b^\nu)^2), 
 \\
\hat{P}_y &=  \frac12 \xi (b^t - b^z) (-b^y h_{+,tz}
+b^x h_{\times,tz} )  
+\mathcal{O}((\xi b^\mu b^\nu)^2), 
 \\
\hat{P}_b &= \frac{1}{2} \xi (b_x^2-b_y^2) h_{+,tt}  
+\xi b^x b^y h_{\times,tt}  
+\mathcal{O}((\xi b^\mu b^\nu)^2) ,
 \\
\hat{P}_l &= - \frac12 \xi^2(b^t-b^z)^2 
[(b_x^2-b_y^2) h_{+,tt}+2b^xb^y h_{\times,tt} ] \nonumber \\
& \hspace{0.5cm} +\mathcal{O}((\xi b^\mu b^\nu)^3) .
\end{align}
\end{subequations}
Four extra polarizations exist and they all depend on the plus mode and the cross mode. The breathing,  vector-x,  and  vector-y modes are suppressed by $\xi b^\mu b^\nu$, while the longitudinal mode is suppressed by $(\xi b^\mu b^\nu)^2$.
Notice that $\Omega$, $\Psi$, $\Sigma_x$, and $\Sigma_y$ also depend on $h_+$ and $h_\times$, but they do not affect the geodesic deviation equation, thus are not shown explicitly here.
We also want to emphasize that the results above are invalid when $b_x^2+b_y^2=0$ or $b^t=b^z$, since during the elimination process, $b_x^2+b_y^2$ and $b^t-b^z$ show up in the denominator.

The dispersion relation of GWs to leading order is given by
\begin{align}
\omega
=& \left( 1 +\frac{1}{2}\xi (b^t-b^z)^2 \right)k   +\mathcal{O}((\xi b^\mu b^\nu)^2) .
\end{align}
The first multi-messenger observations of the binary neutron star merger, i.e.\ GW170817 and GRB\,170817A, placed a tight constraint on the velocity of the tensorial GWs
\cite{LIGOScientific:2017zic}:
\begin{align}
-3\times 10^{-15}    \leq \frac{v_{\text{GW}} -v_{\text{EM}} }{v_{\text{EM}}} \leq +7\times 10^{-16} ,
\end{align}
which implies
\begin{align}
-6\times 10^{-15}  \leq  \xi (b^t-b^z)^2 
\leq +1.4\times 10^{-15} .
\end{align}

For the second solution, the wave equations for the tensor modes are given by
\begin{subequations}
\begin{align}
&-[2-2\xi b_t^2+\xi(2-\rho) b_s^2 ] h_{+,tt} 
+2\xi \rho b^t b^z  h_{+,tz}
\nonumber \\
&+[2-2\xi b_t^2 +2\xi b_s^2+ \xi\rho(b_t^2-b_x^2-b_y^2) ] h_{+,zz} =0 ,
\\
&-[2-2\xi b_t^2+\xi(2-\rho)b_s^2 ] h_{\times,tt} 
+2\xi \rho b^t b^z  h_{\times,tz}
\nonumber \\
&+[2-2\xi b_t^2 +2\xi b_s^2 +\xi\rho(b_t^2-b_x^2-b_y^2) ] h_{\times,zz} =0 .
\end{align}
\end{subequations}
The other perturbations are also not independent and are given by
\begin{subequations}
\begin{align}
\label{s2th}
\Theta &= \frac{(b_x^2-b_y^2 )h_+  +2b^x b^y h_\times}{b_x^2+b_y^2}  , 
 \\
\label{s2phi}
\Phi &= \frac{(b_x^2-b_y^2 )h_+  +2b^x b^y h_\times}{2(b_x^2+b_y^2)}  
+\mathcal{O}(\xi b^\mu b^\nu)  ,
\\
\label{s2xx}
\Xi_x &= -\frac{ (b^t - b^z)(b^x h_+ +b^y h_\times) }{ (b_x^2 +b_y^2 ) }
 +\mathcal{O}(\xi b^\mu b^\nu) ,
\\
\label{s2xy}
\Xi_y &= -\frac{ (b^t - b^z)(-b^y h_+ +b^x h_\times) }{ (b_x^2 +b_y^2 ) }
+\mathcal{O}(\xi b^\mu b^\nu).
\end{align}
\end{subequations}
We surprisingly find that the extra polarizations are not suppressed, but have the same order of amplitude as the tensor polarizations. 
Take $\Theta$ as an example, if $b_x^2 \approx b_y^2 $, then we have $\hat{P}_b \approx \hat{P}_\times$; if $b_x^2 \gg b_y^2 $ or $b_x^2 \ll b_y^2$, then we have $\hat{P}_b \approx \mp \hat{P}_+$. 
If the breathing mode has the same order of amplitude as the plus mode or cross mode, it will distort the waveform recorded in the interferometric detectors, and we should have found the existence of this mode from  real GW data easily. 
On one hand, the residual test conducted in Ref.~\cite{LIGOScientific:2021sio} shows the consistency of the signals in the data with GR.
On the other hand, it is indicated in Ref.~\cite{Takeda:2021hgo} that the scalar-mode amplitude should be at least one order of magnitude smaller than the tensor-mode amplitude.
We expect that the amplitude of the extra polarizations is much smaller compared to the tensor polarizations, thus we do not consider this set of solutions.

\subsection{Case (II): $b^x=b^y=0$ and $b^t\neq b^z$ }
\label{bxby0}

If $b^x=b^y=0$, 
we can get the wave equations for the plus mode and the cross mode directly from Eq. \eqref{bbbxmy} and Eq. \eqref{bbbxy}, which are the same as Eq. \eqref{bumso1p} and Eq. \eqref{bumso1c}.
As for the vector perturbations,
combining Eqs.~\eqref{bbbvx}-\eqref{bbbyz}, we can obtain the following wave equations:
\begin{widetext}
\begin{subequations}
\begin{align}
 -[ 2-2\xi b_t^2 +\xi(2-\rho)b_z^2 ]\Sigma_{x,tt}
+2\xi \rho b^t b^z  \Sigma_{x,tz}
+[ 2-\xi (2-\rho)b_t^2 +2\xi b_z^2 ]\Sigma_{x,zz} = 0 , 
\\
 -[ 2-2\xi b_t^2 +\xi (2-\rho) b_z^2 ]\Sigma_{y,tt}
+2\xi \rho b^t b^z  \Sigma_{y,tz}
+[ 2-\xi (2-\rho) b_t^2 +2\xi b_z^2 ]\Sigma_{y,zz} = 0 , 
\\
\Xi_x + \xi \frac{2b^t-b^z \sqrt{[2-2\xi(b_t^2-b_z^2)]
[2 -\xi(2-\rho)(b_t^2-b_z^2) ]   } }
{2+\xi b_z^2[4-\rho -\xi(2-\rho)(b_t^2-b_z^2)]}\Sigma_x = 0, 
\\
\Xi_y + \xi \frac{2b^t-b^z \sqrt{[2-2\xi(b_t^2-b_z^2)]
[2 -\xi(2-\rho)(b_t^2-b_z^2) ]   } }
{2+\xi b_z^2[4-\rho -\xi(2-\rho)(b_t^2-b_z^2)]}\Sigma_y = 0 .
\end{align}
\end{subequations}
\end{widetext}
As we can see, only two of the four vector perturbations $\Sigma_x$, $\Sigma_y$, $\Xi_x$ and $\Xi_y$ are independent. They will activate the vector-x mode and the vector-y mode. If $\rho=2$, then they have the same dispersion relation as the tensor modes.
Last but not least, we can get the equations for the scalar perturbations using Eq. \eqref{bbbxpy} and Eq. \eqref{bbbc}:
\begin{subequations}
\begin{align}
0 &= -[3-3\xi b_t^2 +2\xi(2-\rho)b_z^2]\Theta_{,tt}
-2\xi (1-2\rho) b^t b^z \Theta_{,tz}
\nonumber \\
& \hspace{0.5cm} +[3 -2\xi(2-\rho) b_t^2 +3\xi b_z^2]\Theta_{,zz} 
\nonumber \\
& \hspace{0.5cm} +\frac{8\lambda}{\xi\rho}
(1-\xi b_t^2+\xi b_z^2)^2\Theta  ,   
\\
\Phi_{,zz} &=\frac{ -\xi(1-2\rho)b_z^2\Theta_{,tt}
-2\xi (1-2\rho) b^t b^z \Theta_{,tz} }
{2 -2\xi b_t^2 +2\xi b_z^2} 
\nonumber \\
& \hspace{0.5cm} 
+ \frac{[1 -2\xi(1-\rho) b_t^2 +\xi b_z^2] \Theta_{,zz}}{2 -2\xi b_t^2 +2\xi b_z^2}  .
\end{align}
\end{subequations}
There is a $\lambda$-term in the wave equation, which acts as a mass term. Besides, $\Omega$ and $\Psi$ also depend on $\Theta$. There is only one independent DOF for scalar perturbations, which will activate a mixture of the breathing mode and the longitudinal mode.

It reduces to the results for a purely timelike background or a purely spacelike background when we set 
$b^z=0$ or $b^t=0$ respectively.
Notice that, differently from the results we obtained in the previous case, the vector and scalar perturbations do not depend on the tensor perturbations and they propagate at different speeds, i.e.\ they are independent modes.

\subsection{ Case (III): $b^t=b^z$}
\label{bt=bz}

If $b^t =b^z$, using a similar analysis, we find that $\Xi_x$, $\Xi_y$, $\Theta$, and $\Phi$ do not propagate, while the two tensor perturbations, $h_+$ and $h_\times$ propagate at the speed of light. Thus, there are only two DOFs in GWs in this case.

\subsection{Preferred Coordinate System}
\label{preframe}

Our previous calculations adopt the GW coordinate system $(t,x,y,z)$ in which GWs propagate along the $z$ axis, which simplifies the calculation.
In this subsection, we will explain the results in another coordinate system $(t,x',y',z')$  where $b^{\prime\,\mu}=(b^t,0,0,b_s)$, to help us understand the effects of Lorentz violation better.
We will call the coordinates $(t,x',y',z')$ the preferred coordinate system of the background $b^\mu$, since the spatial part of the bumblebee background field lies along the $z'$ axis. The two coordinate systems can be related by an Euler rotation $E_{ij}(\alpha,\beta,\phi)$, which is defined via
\begin{subequations}
\begin{align}
k'_i=E_{ij}(\alpha,\beta,\phi) k_j , \\
b_i=E^{-1}_{ij}(\alpha,\beta,\phi) b'_j ,
\end{align}
\end{subequations}
where
\begin{widetext}
\begin{align}
E_{ij}= \left(
\begin{array}{ccc}
\cos\beta\cos\alpha\cos\phi -\sin\alpha\sin\phi   &   -\cos\beta\cos\alpha\sin\phi -\sin\alpha\cos\phi  & \sin\beta\cos\alpha          \\
\cos\beta\sin\alpha\cos\phi +\cos\alpha\sin\phi   &     -\cos\beta\sin\alpha\sin\phi +\cos\alpha\cos\phi  &
\sin\beta\sin\alpha \\
 -\sin\beta\cos\phi    & \sin\beta\sin\phi & \cos\beta
\end{array}
\right) ,
\end{align}
\end{widetext}
and $k_i$ and $k'_i$ are the spatial components of the wave vector in the GW coordinate system and the preferred coordinate system respectively. Their unit vectors are denoted as  
$\hat{\bm{k}}=(0,0,1)$ and $\hat{\bm{k}}'=(\sin\beta\cos\alpha,\sin\beta\sin\alpha,\cos\beta)$.
We also have $\bm{b}'=(0,0,b_s)$ and 
$\bm{b}=(b_x,b_y,b_z)=(-b_s\sin\beta\cos\phi,b_s\sin\beta\sin\phi,b_s\cos\beta)$.
From the GW coordinate system to the preferred coordinate system,
the scalar perturbations remain unchanged, while the vector and tensor perturbations transform as follows,
\begin{subequations}
\begin{align}
\Xi'_i &= E_{ij} \Xi_j ,
\\
\Sigma'_i &= E_{ij} \Sigma_j ,
\\
{h'}_{ij}^{\text{TT}} &= E_{ik}E_{jl} h_{kl}^{\text{TT}} .
\end{align}
\end{subequations}

Here, we take the $xy$ component of Eq. \eqref{Bumgfieldo} as an example. In the preferred coordinate system, it turns out to be
\begin{widetext}
\begin{align}
&[(3+\cos(2\beta))\cos(2\phi)\sin(2\alpha) +4\cos\beta\sin(2\phi)\cos(2\alpha) ]
\nonumber \\
& \hspace{0.5cm} \times [ (-1+\xi b_t^2) \partial^2_t h_+(t,n)
+2\xi b^t b_s  \cos\beta \partial_t \partial_{n} h_+(t,n) 
+ (1+\xi b_s^2 \cos^2\beta) \partial^2_{n} h_+(t,n) ]
\nonumber \\
+&  [-(3+\cos(2\beta))\sin(2\phi)\sin(2\alpha) +4\cos\beta\cos(2\phi)\cos(2\alpha) ]
\nonumber \\
& \hspace{0.5cm} \times [ (-1+\xi b_t^2) \partial^2_t h_\times(t,n)
+2\xi b^t b_s  \cos\beta \partial_t \partial_{n} h_\times(t,n) 
+ (1+\xi b_s^2 \cos^2\beta) \partial^2_{n} h_\times(t,n) ]
\nonumber \\
+& \sin\beta f(\Xi_x, \Xi_y, \Sigma_x, \Sigma_y, \Theta, \Phi, \Psi)
=0 ,
\end{align}
\end{widetext}
where $n=z'\cos\beta +x'\sin\beta\cos\alpha +y' \sin\beta\cos\alpha$, and $f$ is a linear combination of the spacetime derivatives of the vector and scalar perturbations, which we will not show explicitly here. 
If $\beta=0$ or $\pi$, then $\sin\beta=0$ so that only the tensor perturbations are left in this equation. We can get similar results for the other components of the field equations, indicating that the tensor, vector and scalar perturbations decouple from each other. This is exactly the Case (II) which we discussed in Sec.~\ref{bxby0}.

If $|b^t|<b_s$, i.e.\ the bumblebee background is spacelike, there will be an extra special direction corresponding to $\beta=\beta^\star=\arccos(b^t/b_s)$. That is the Case (III) which we discussed in Sec.~\ref{bt=bz}.

For $\beta \notin \{ 0, \pi, \beta^\star \}$, it is the Case (I) where the vector and scalar perturbations couple to the tensor perturbations and the dispersion relation becomes
\begin{align}
\omega=\frac{-\xi b^t b_s \cos\beta
+\sqrt{1-\xi b_t^2 +\xi b_s^2 \cos^2\beta } }
{1-\xi b_t^2 }k ,
\end{align}
indicating that the propagation speed of GWs depends on the angle $\beta$ between $\hat{\bm{k}}$ and $\bhat$.
Besides, the amplitude of the extra polarizations also depends on $\beta$.
This is the anisotropy induced by the background field.
The dispersion relation and the polarization content are independent of the azimuthal angle $\alpha$. It is not surprising, since in the preferred coordinate system, the bumblebee background is axisymmetric around the $z'$ axis.

If $|b^t|>b_s>0$, then we have $b^\mu b_\mu<0$, and the bumblebee background is timelike. For observers in a purely timelike background, the rotation symmetry is preserved and there are always five DOFs in GWs, regardless of which direction GWs come from. 
But when the spatial component shows up, there are preferred spatial directions causing anisotropy in GWs. Notice that even when $b^\mu$ only has the temporal component for one observer, another observer with a relative velocity sees spatial components of $b^\mu$ and therefore can perceive a different polarization content of GWs. This simply reflects the violation of the boost symmetry in the bumblebee model.
However, it does not make a significant difference for the polarization content in a spacelike background, whether the temporal component is zero or not.
We summarize our results in Table \ref{bumsum}.

In the no-coupling limit, i.e. $\xi\to 0$, only
the tensor modes of GWs can propagate, and they remain unaffected by the vector background. This result is consistent with that in Ref.~\cite{Bluhm:2007bd}.

\begin{table}[h]
\renewcommand\arraystretch{1.3}
\centering
\caption{Summary for polarizations of GWs in the bumblebee model.  Here we denote the bumblebee background as $b^\mu=(b^t,b_s\bhat)$, where $\bhat$ is a unit vector representing the direction of the spatial part of $b^\mu$; $\hat{\bm{k}}$ is the propagation direction of GWs and $\beta$ is the angle between $\hat{\bm{k}}$ and $\bhat$. 
The column ``Case'' corresponds to the classification in Sec.~\ref{bumblepolar}, in the GW coordinate system.
In the last column,
``TVS coupled'' means that the tensor, vector, and scalar modes are coupled to each other;
``2T+2V+1S'' means that there are independently two tensor modes, two vector modes, and one scalar mode in polarizations of GWs. ``2T'' means that there are only two tensor modes.
}
\label{bumsum}
\begin{tabular*}{\hsize}{@{}@{\extracolsep{\fill}}lcccccc@{}}
\toprule
\multicolumn{2}{l}{ Bumblebee field} & $\beta$ &  Case
& DOFs & Polarization 
\\
\hline
Timelike & $b_s=0 $  & -- &  (II)
& 5  & 2T+2V+1S \\
\cline{2-6}
       & $b_s\neq0$ &  $\beta \in \{0, \pi\}$ &   (II)
       & 5 & 2T+2V+1S \\
       &  &  $\beta \notin \{0, \pi\}$ &  (I)
       & 2 & TVS coupled \\
\hline
Spacelike  &  & $\beta \in \{0, \pi\}$ &  (II)
                 & 5 & 2T+2V+1S \\
       &  & $\beta^\star=\arccos(b^t/b_s)$ &  (III)
                  & 2 & 2T \\
         &  & $\beta \notin \{0, \pi, \beta^\star\}$ &  (I)
                  & 2 & TVS coupled \\
\hline
\end{tabular*}
\end{table}

\section{Bumblebee-like SME model}
\label{comparison}

The SME framework contains the Largrangian densities for both the
Standard Model and GR, along with all scalar terms involving operators for Lorentz violation and CPT violation, offering a general parameterization of Lorentz and CPT violation \cite{Colladay:1996iz, Colladay:1998fq,Kostelecky:2003fs,
Bailey:2006fd}. 
Bumblebee gravity model is only one of the specific and explicit  models in the SME framework to illustrate Lorentz violation effects in the gravity sector. 
In this section, we consider the propagation of GWs in the linearized regime in the SME framework, to compare with the results in the bumblebee gravity model.
To begin with, the action in the SME framework is given by \cite{Bailey:2006fd}:
\begin{equation}
S=\int d^4x \sqrt{-g} \frac{R}{2\kappa} +S_{\text{LV}} +S'  ,
\end{equation}
where $S_{\text{LV}} $ contains the leading Lorentz-violating couplings as follows
\begin{equation}
\label{SLV}
S_{\text{LV}} =\frac{1}{2\kappa} \int d^4x \sqrt{-g} (-u R +s^{\mu\nu}R^{\rm T}_{\mu\nu} +t^{\kappa\lambda\mu\nu}C_{\kappa\lambda\mu\nu} ).
\end{equation}
Here, $u$, $s^{\mu\nu}$, and $t^{\kappa\lambda\mu\nu}$ are the fields inducing Lorentz violation,
$R^{\rm T}_{\mu\nu}$ is the trace-free Ricci tensor, and $C_{\kappa\lambda\mu\nu}$ is the Weyl tensor. The tensor field $s^{\mu\nu}$ is taken to be symmetric and traceless. The tensor field $t^{\kappa\lambda\mu\nu}$ inherits the symmetries of the Riemann tensor, and its trace and partial traces vanish. The action $S'$ includes dynamics not only for conventional matter but also for the Lorentz-violating fields $u$, $s^{\mu\nu}$ and $t^{\kappa\lambda\mu\nu}$.

The variation with respect to the metric yields the field equation in vacuum:
\begin{equation}
G_{\mu\nu}=(T^{Rstu})_{\mu\nu} +\kappa (T^{stu})_{\mu\nu},
\end{equation}
where
\begin{align}
\label{TRstu}
(T^{Rstu})_{\mu\nu}=& -\frac{1}{2}D_\mu D_\nu u -\frac{1}{2}D_\nu D_\mu u
+g_{\mu\nu}D^\alpha D_\alpha u 
\nonumber \\
& +u G_{\mu\nu}
+\frac{1}{2}D^\alpha D_\mu s_{\alpha\nu} +\frac{1}{2}D^\alpha D_\nu s_{\alpha\mu} 
\nonumber \\
& -\frac{1}{2}D^2 s_{\mu\nu} -\frac{1}{2}g_{\mu\nu}D_\alpha D_\beta s^{\alpha\beta} 
+\frac{1}{2}g_{\mu\nu}s^{\alpha\beta}R_{\alpha\beta} 
\nonumber \\
& -D^\alpha D^\beta t_{\mu\alpha\nu\beta}
-D^\alpha D^\beta t_{\nu\alpha\mu\beta} 
\nonumber \\
& +\frac{1}{2}{t^{\alpha\beta\gamma}}_{\mu}R_{\alpha\beta\gamma\nu}
+\frac{1}{2}{t^{\alpha\beta\gamma}}_{\nu}R_{\alpha\beta\gamma\mu}
\nonumber \\
& +\frac{1}{2}g_{\mu\nu}t^{\alpha\beta\gamma\delta}R_{\alpha\beta\gamma\delta} ,
\end{align}
which comes from the variation of $S_\text{LV}$ with respect to the metric
and $ (T^{stu})_{\mu\nu}$ is the energy-momentum tensor contributed from the dynamics of $u$, $s^{\mu\nu}$, and $t^{\kappa\lambda\mu\nu}$.
Although the form of $S'$ is not given explicitly, the effective linearized field equation 
\begin{equation}
\label{SMEfieldeq}
\begin{aligned}
R_{\mu\nu}=a_1 \bar{u} R_{\mu\nu}  + a_2 \bigg( & \frac{1}{2}\eta_{\mu\nu}\bar{s}^{\alpha\beta} R_{\alpha\beta}-2\bar{s}^\alpha_{\pt\alpha(\mu}R_{\alpha\nu)}
\\
& +\frac{1}{2}\bar{s}_{\mu\nu}R +\bar{s}^{\alpha\beta} R_{\alpha\mu\nu\beta} \bigg)
\end{aligned}
\end{equation}
was derived by \citet{Bailey:2006fd} under several assumptions. In Eq.~\eqref{SMEfieldeq}, $\bar{u}$ and $\bar{s}^{\mu\nu}$ are the vacuum expectation values of the Lorentz-violating fields, which are assumed to be constant. Besides, $a_1$ and $a_2$ are coefficients, which vary with the specific theory.
One of the important assumptions made by \citet{Bailey:2006fd} is the assumption (v) that all the terms in $(T^{Rstu})_{\mu\nu}$ and $(T^{stu})_{\mu\nu}$ can be constructed from the linear combinations of the vacuum values $\eta_{\mu\nu}$, $\bar{u}$, $\bar{s}^{\mu\nu}$, and the two spacetime derivatives of $h_{\mu\nu}$.
Thus, it represents for a general subclass of models in the SME framework.
As was pointed out in Ref.~\cite{Bailey:2006fd}, the bumblebee model is an explicit model that weakly violates the assumption (iv) therein, namely, that
the independently conserved piece of the trace-reversed energy momentum tensor does not vanish.
Thus, it is interesting to compare the propagation of GWs in the bumblebee model and the linearized SME model in Ref.~\cite{Bailey:2006fd}.  

Up to the linear order, the coupling term $\kappa T^\xi_{\mu\nu}$ in Eq.~\eqref{tensoreq} can be reproduced by $(T^{Rstu})_{\mu\nu}$ in the SME framework, i.e. Eq.~\eqref{TRstu}, by making the following identifications
\begin{subequations}
\begin{align}
\bar{u} &=\frac{1}{4}\xi b^\alpha b_\alpha , 
\\
\tilde{u} &= \frac{1}{2}\xi \left( b_\alpha \tilde{B}^
\alpha +\frac{1}{2}b^\alpha b^\beta h_{\alpha\beta} \right) ,  
\\
\bar{s}^{\mu\nu} &= \xi \left(b^\mu b^\nu -\frac{1}{4}\eta^{\mu\nu}b^\alpha b_\alpha  \right) ,
\\
\tilde{s}^{\mu\nu} &= \xi \big(
b^\mu \tilde{B}^\nu 
+b^\nu \tilde{B}^\mu 
-\frac{1}{2}b_\alpha \tilde{B}^\alpha
-\frac{1}{4} \eta^{\mu\nu}b^\alpha b^\beta
\nonumber \\
& \hspace{1cm} +\frac{1}{4}h^{\mu\nu}b^\alpha b_\alpha \big)  h_{\alpha\beta} , 
\\
\bar{t}^{\alpha\beta\gamma\delta} &=0, 
\\
\tilde{t} ^{\alpha\beta\gamma\delta} &=0,
\end{align}
\end{subequations}
where $\tilde{u}$, $\tilde{s}^{\mu\nu}$, and $\tilde{t}^{\alpha\beta\gamma\delta}$ are the perturbations of these fields around their vacuum expectation values.
Now we substitute these identifications into Eq. \eqref{SMEfieldeq}, and follow Ref.~\cite{Bailey:2006fd} to  adopt $a_1=-3$, $a_2=1$. Then the field equation can be rewritten as
\begin{align}
\label{bSME}
G_{\mu\nu}=\xi \bigg( & \frac12 b_\mu b_\nu R +\eta_{\mu\nu} b^\alpha b^\beta R_{\alpha\beta} 
-b^\alpha b_\mu R_{\alpha\nu} \nonumber \\
& -b^\alpha b_\nu R_{\alpha\mu} 
-b^\alpha b^\beta R_{\alpha\mu\beta\nu} \bigg) ,
\end{align}
and we call it the bumblebee-like SME model.
In this model, the details about the derivation of the wave solutions are given in Appendix~\ref{bsmeappendix}. 
The wave solutions are found to be exactly the same as 
the first set of solutions in the bumblebee model in the Case (I), i.e. Eqs.~\eqref{bumso1p} and \eqref{bumso1c}, and Eqs.~\eqref{s1th} to \eqref{s1xiy}. 

As is shown in Appendix~\ref{bsmeappendix}, during the calculation, the denominator is always nonzero. Thus, there are always two DOFs in GWs in the bumblebee-like SME model, which is different from the original bumblebee model. 
According to Eqs.~\eqref{s1th} to \eqref{s1xiy}, 
if $b^x=b^y=0$,
all the extra polarizations vanish, which means that there are only the plus and cross modes when $\beta=0$ or $\pi$.
In the original bumblebee model, when GWs propagate along $\bhat$, the vector and scalar perturbations decouple from the tensor perturbations and then become independent. The vector/scalar part of $h_{\mu\nu}$ has connections with the vector/scalar part of $\tilde{B}_\mu$, which produces the vector/scalar polarization.
For a spacelike background, when $b^t=b^z$, or equivalently $\beta=\beta^\star$, one has $\hat{P}_x=\hat{P}_y=\hat{P}_l=0$.
Only plus mode, cross mode, and breathing mode are nonzero, and the breathing mode is not independent.
These perturbations propagate at the speed of light,  as is indicated by Eq.~\eqref{c1tensordp}. 
For $\beta\notin \{ 0, \pi, \beta^\star \}$, all extra polarizations show up, but the total number of DOFs is two.

To conclude, for a purely timelike background, there are five independent DOFs of GWs in the bumblebee model while only two independent polarizations exist in the bumblebbe-like SME model. 
However, when the spatial components of $b^\mu$ exist, the propagation and polarizations of GWs are the same in these two models, except when GWs come from  special directions, i.e. $\beta \in \{ 0, \pi, \beta^\star \}$.
The results are summarized in Table \ref{smesum}.

\begin{table}
\renewcommand\arraystretch{1.3}
\centering
\caption{Similar to Table~\ref{bumsum}, for the bumblebee-like SME model. ``TS coupled'' means that the tensor and scalar modes are coupled to each other.
}
\label{smesum}
\begin{tabular*}{\hsize}{@{}@{\extracolsep{\fill}}lccccl@{}}
\toprule
\multicolumn{2}{l}{Bumblebee field} & $\beta$ & DOFs & Polarization
\\
\hline
Timelike & $b_s=0 $  & -- & 2  & 2T \\
\cline{2-5}
       & $b_s\neq0$ &  $\beta \in \{0, \pi \}$ & 2 & 2T \\
       &  &  $\beta \notin \{0, \pi \}$ & 2 & TVS coupled \\
\hline
Spacelike &  &  $\beta \in \{0, \pi \}$ & 2 & 2T \\
         &  & $\beta^\star=\arccos(b^t/b_s)$ & 2 & TS coupled \\
         &  &  $\beta \notin \{0, \pi, \beta^\star \}$ & 2 & TVS coupled \\
\hline
\end{tabular*}
\end{table}

Notice that, it is assumed that all the contributions from $(T^{Rstu})_{\mu\nu}$ and $(T^{stu})_{\mu\nu}$ can be constructed from the linear combination of  two spacetime derivatives of $h_{\mu\nu}$ when deriving the linearized field equation \cite{Bailey:2006fd}.
In the bumblebee gravity model, some gauge invariants are constructed by the combination of the vector field perturbation and the metric perturbation, which will not directly reduce to the two spacetime derivatives of the metric perturbation.
In the metric field equation, terms like $\Sigma_{x,tz}$ and $\Sigma_{y,tz}$ cannot be eliminated via the vector field equation.
Thus, we anticipate that it is the violation of assumption (v) in Ref.~\cite{Bailey:2006fd} that makes the bumblebee-like SME model different from the original bumblebee gravity model.

\section{Discussion and Conclusion}
\label{discussion}

We introduced a vector field called bumblebee field to induce Lorentz violation, and the effects of a preferred spatial direction defined by the vector field are manifested in the anisotropy.
The interaction between the background vector field and the electromagnetic field causes anisotropy and modifies dispersion relation for photons,
and these Lorentz-violating effects have been tightly constrained by current astrophysical and cosmological observations \cite{Kostelecky:2008be,Planck:2018vyg}.
While we concentrate on the interaction between the Lorentz-violating field and gravity in this paper. 
The search for the anisotropy, dispersion, and birefringence of GWs was performed in GW observations \cite{Kostelecky:2016kfm,Shao:2020shv,Wang:2021gqm,Zhao:2022pun,Niu:2022yhr,Haegel:2022ymk}, and the results show that these Lorentz-violating effects are sufficiently small. Thus, we have $|\xi b^\mu b^\nu| \ll 1$ in the bumblebee gravity model. 
In addition to the modifications on the propagation of GWs, extra polarizations of GWs are activated due to Lorentz violation. 
However, the amplitude of the extra polarizations is suppressed by $|\xi b^\mu b^\nu|$ in the bumblebee gravity model, making them hard to detect. 
But our theoretical study finds some novel phenomena, which will provide deeper understanding on the nature of the polarizations of GWs in the future.

In this paper, we use gauge invariants to investigate the polarization content of GWs in the bumblebee gravity model. 
There are five DOFs of the propagating modes in total in a purely timelike background. 
In addition to the plus mode and cross mode, there are two vector modes and one scalar mode, which is similar to the polarization content in another vector-tensor theory, the Einstein-\ae{}ther theory.
The scalar perturbation activates a mixture of the breathing mode and the longitudinal mode.
The tensor modes and vector modes are massless and the scalar mode is massive due to the potential in the action~(\ref{bumblebeeaction}). 

For a background containing a spatial component $b_s$, it defines a preferred spatial direction $\bhat$ which breaks rotational symmetry.
We are the first to analyze the polarizations of GWs in such a spacelike vector background.
We denote $\beta$ as the angle between the propagation direction $\hat{\bm{k}}$ of GWs and $\bhat$.
When $\beta=0$ or $\pi$,  there are five independent DOFs. The polarization content is similar to the purely timelike case. Tensor, vector and scalar perturbations propagate at different speeds.
For a spacelike background, when we have $|b^t|<b_s$, then $\beta^\star=\arccos(b^t/b_s)$ defines another special direction.
There are only the plus and cross modes for GWs propagating along the direction $\beta = \beta^\star$, and the speed is $c$.

For a general $\beta \notin \{0, \pi, \beta^\star \}$, all the tensor, vector,
and scalar perturbations couple to each other.  Technically, we introduce
higher-order derivatives to separate the perturbations, and obtain two sets of
possible solutions. One of the solutions predicts that the extra polarizations
have the amplitude as large as the tensor polarizations, thus it is excluded by
the current observation.  For the remaining viable solution, the amplitude of
the extra polarizations is suppressed by $\xi b^\mu b^\nu$.  Although all six
polarizations are present, the four extra polarizations depend on the tensor
modes. In other words, the tensor perturbations can activate a mixture of all
the six  polarizations simultaneously. The results are shown in Table
\ref{bumsum}.  In general, the polarizations are defined by the geodesic
deviation equations. They are defined kinematically, thus they are not
equivalent to the dynamical DOFs.  To our best knowledge, the coupling of the
tensor, vector and scalar perturbations is a new phenomenon unexplored before. 
Our study provides more theoretical
understanding of the interaction between the Lorentz-violating vector field and
the metric.

We also analyze the polarizations in the SME framework in the linearized regime.
If $s^{\mu\nu}$ and $u$ are represented by a vector field $B^\mu$, then we call it the bumblebee-like SME model.
For the bumblebee-like SME model, there are only the plus and cross modes in a purely timelike background. 
If $\beta=0$ or $\pi$, only two tensor modes are present. If $\beta=\beta^\star$, there is an additional breathing polarization, but it is not independent.
For a general $\beta \notin \{0, \pi, \beta^\star \}$, the wave solutions are identical to the first sets of solutions in our Case (I) in the bumblebee model (see Sec.~\ref{bumgeneral}). The results are shown in Table \ref{smesum}.
Unlike in the original bumblebee gravity model, there are always two DOFs of GWs in the bumblebee-like SME model, no matter along which direction GWs come. 
We propose the conjecture that the original bumblebee model is equivalent to some effective SME model with higher-order derivatives in the linearized regime. It is out of the scope of this paper, and we leave it to a future study.

\acknowledgments 
We acknowledge the anonymous referees for the useful comments, which help us improve our paper.
We thank V.\ Alan Kosteleck\'y, Yungui Gong, Shaoqi Hou and Marco Schreck for helpful comments and discussions.
This work was supported by the National Natural Science Foundation of China (12147120, 11975027, 11991053,  11721303), the China Postdoctoral Science Foundation (2021TQ0018),
the National SKA Program of China (2020SKA0120300), 
the Max Planck Partner Group Program funded by the Max Planck Society, 
and the High-Performance Computing Platform of Peking University.
RX is supported by the Boya Postdoctoral Fellowship at Peking University.

\appendix

\section{Gauge invariants}
\label{gi}

The Riemann tensor, Ricci tensor, and Ricci scalar can be expressed in terms of the gauge invariants as follows
\begin{align}
R & =3\Theta_{,tt} -2\nabla^2 \Theta -2\nabla^2\Phi ,
\\
R_{tt} & =\frac{1}{2} (-3\Theta_{,tt}+2\nabla^2 \Phi ) ,
\\
R_{ti} & =\frac{1}{2}(-2\Theta_{,ti}-\nabla^2 \Xi_i ) ,
\\
R_{ij} & = \frac{1}{2}(- \delta_{ij}\Box \Theta  -\Theta_{,ij}  -2\Phi_{,ij}   -\Xi_{i,tj} -\Xi_{j,ti} -\Box h_{ij}^{\text{TT}}) ,
\\
R_{titj} & =\frac{1}{2}(-\delta_{ij}\Theta_{,tt}+2\Phi_{,ij}+ \Xi_{i,tj}+\Xi_{j,ti} -h_{ij,tt}^{\text{TT}} ) ,
\\
R_{tijk} & = \frac{1}{2}(
\delta_{ij}\Theta_{,tk}-\delta_{ik}\Theta_{,tj}
+\Xi_{k,ij}-\Xi_{j,ik}
\nonumber \\
& \hspace{1cm} +h^{\text{TT}}_{ij,tk}-h^{\text{TT}}_{ik,tj}  ) ,
\\
R_{ijkl} & =\frac{1}{2}( \delta_{il}\Theta_{,jk}+\delta_{jk}\Theta_{,il}
-\delta_{ik}\Theta_{,jl}-\delta_{jl}\Theta_{,ik} 
\nonumber \\
& \hspace{1cm} +h_{il,jk}^{\text{TT}}+ h_{jk,il}^{\text{TT}}
-h_{ik,jl}^{\text{TT}}-h_{jl,ik}^{\text{TT}} ) .
\end{align}
The quantities in Eq. \eqref{Bumvfieldo} and Eq. \eqref{Bumgfieldo} can be written in terms of the gauge invariants as follows
\begin{widetext}
\begin{align}
V' & =\lambda( -2b^t \Omega  +2b^i\Sigma_i +2b^i \Psi_{,i} -2b_t^2 \Phi +2b^t b^i \Xi_{i} +b^i b^j (h_{ij}^{\text{TT}} +\delta_{ij}\Theta) )  \ ,
\\
C_t & = -\nabla^2 \Omega   -\nabla^2 \Psi_{,t} -2b^t \nabla^2 \Phi +b^i\nabla^2 \Xi_i -b^i \Theta_{,ti} \ ,
\\
C_i  & =  -\Omega_{,ti} -\Psi_{,tti} +\Box \Sigma_i 
 +b^t( \Box \Xi_{i} -2\Phi_{,ti} ) +b^i \Box \Theta 
 + b^j ( \Box h_{ij}^{\text{TT}} +\Xi_{j,ti} -\Theta_{,ij} ) \ ,
\\
\bar{A}_{tt} & = -2 b^t \Omega_{,tt} -2b^i \Omega_{,ti} -2b_t^2 \Phi_{,tt} -4b^t b^i \Phi_{,ti} -2b^i b^j \Phi_{,ij} \ ,
\\
\bar{A}_{ti} & = b^t(- \Omega_{,ti}  + \Sigma_{i,tt} + \Psi_{,tti}) +b^j (-\Omega_{,ij} +\Sigma_{i,tj} + \Psi_{,tij} )
+b_t^2 \Xi_{i,tt} +2 b^t b^j \Xi_{i,tj} +b^j b^k \Xi_{i,jk} \ ,
\\
\bar{A}_{ij} & = b^t (\Sigma_{i,tj} + \Sigma_{j,ti} + 2\Psi_{,tij}) +b^k(\Sigma_{i,jk}+\Sigma_{j,ik} +2\Psi_{,ijk}) 
\nonumber \\
& \hspace{0.5cm} + b_t^2 (h_{ij,tt}^{\text{TT}} +\delta_{ij}\Theta_{,tt} ) +2 b^t b^k (h_{ij,tk}^{\text{TT}} +\delta_{ij}\Theta_{,tk} ) +b^k b^l  (h_{ij,kl}^{\text{TT}} +\delta_{ij}\Theta_{,kl} )  \ ,
\\
\bar{D} & = 2b^t \Omega_{,tt}+2b^i\Omega_{,ti} +2b^t \nabla^2\Psi_{,t} +2b^i \nabla^2 \Psi_{,i}  
 +b_t^2(2\Phi_{,tt}+\nabla^2\Phi +\frac{3}{2}\Theta_{,tt} ) +b^t b^i (4\Phi_{,ti}+4\Theta_{,ti}-\nabla^2\Xi_i ) 
\nonumber \\ 
& \hspace{0.5cm}  + b^i b^j ( -\frac{1}{2}\Box h_{ij}^{\text{TT}} -\frac{1}{2}\delta_{ij}\Box \Theta +\frac{5}{2}\Theta_{,ij} +\Phi_{,ij} -\frac{1}{2}\Xi_{i,tj}-\frac{1}{2}\Xi_{j,ti} ) \ .
\end{align}
\end{widetext}

\section{Calculation details in the bumblebee gravity model}
\label{bbbappendix}

To simplify the calculation, we assume that the metric field and the bumblebee field propagate along the $+z$ axis. To obtain the  wave equations for the perturbations separately, we adopt an elimination method.
During the elimination process, we have to be cautious, and check if the denominator is zero.

We use Eq. \eqref{vprime} with $t$, $z$ components of Eq. \eqref{Bumvfieldo} and $tt$, $tz$, $zz$ components of Eq. \eqref{Bumgfieldo} to eliminate $\Omega$ and $\Psi$ first, since they do not affect the geodesic deviation equation.
For $b^z\neq0$, we have
\begin{widetext}
\begin{align} 
\label{elim1}
\Psi_{,ttz} &= \frac{1}{\xi b^z} f_1
(\Omega_{,tt},\Phi_{,tt},\Theta_{,tt}  )  , 
\nonumber  \\
\Psi_{,tzz} &= \frac{1}{\xi b_z^2} f_2 
(\Omega_{,tt},\Phi_{,tt},\Theta_{,tt},\Phi_{,tz},\Theta_{,tz},\Xi_{x,tz},\Xi_{y,tz}  ) ,  
\nonumber \\
\Psi_{,zzz} &= \frac{1}{\xi b_z^3} f_3 (\Omega_{,tt},\Phi_{,tt},\Theta_{,tt},\Phi_{,tz},\Theta_{,tz},\Xi_{x,tz},\Xi_{y,tz},\Phi_{,zz},\Theta_{,zz},\Xi_{x,zz},\Xi_{y,zz}  )   , 
\nonumber \\
\Omega_{,tz} &= \frac{1}{\xi b^z} f_4 (\Omega_{,tt},\Phi_{,tt},\Theta_{,tt},\Phi_{,tz},\Theta_{,tz},\Xi_{x,tz},\Xi_{y,tz}  )      , 
\nonumber \\
\Omega_{,zz} &= \frac{1}{\xi b_z^2} f_5 
(\Omega_{,tt},\Phi_{,tt},\Theta_{,tt},\Phi_{,tz},\Theta_{,tz},\Xi_{x,tz},\Xi_{y,tz},\Phi_{,zz},\Theta_{,zz},\Xi_{x,zz},\Xi_{y,zz}  ) . 
\end{align}
For $b^t\neq0$, we have
\begin{align}
\label{elim2}
\Psi_{,tzz} &= \frac{1}{\xi b^t} g_1 
(\Psi_{,zzz}, \Phi_{,zz}, \Theta_{,zz}, \Xi_{x,zz}, \Xi_{y,zz}, \Theta_{,tz}   ),  
\nonumber \\
\Psi_{,ttz} &= \frac{1}{\xi b_t^2} g_2 
(\Psi_{,zzz}, \Phi_{,zz}, \Theta_{,zz}, \Xi_{x,zz}, \Xi_{y,zz}, \Phi_{,tz},  \Theta_{,tz}, \Xi_{x,tz}, \Xi_{y,tz}, \Theta_{,tt} ),
\nonumber \\
\Omega_{,zz} &= \frac{1}{\xi b^t} g_3  
(\Psi_{,zzz},\Phi_{,zz}, \Theta_{,zz} ) ,
\nonumber \\ 
\Omega_{,tz} &= \frac{1}{\xi b_t^2} g_4 
(\Psi_{,zzz}, \Phi_{,zz}, \Theta_{,zz}, \Xi_{x,zz}, \Xi_{y,zz}, \Phi_{,tz}, \Theta_{,tz} )  ,  
\nonumber  \\
\Omega_{,tt} &= \frac{1}{\xi b_t^3} g_5 
(\Psi_{,zzz}, \Phi_{,zz}, \Theta_{,zz}, \Xi_{x,zz}, \Xi_{y,zz}, \Phi_{,tz}, \Theta_{,tz}, \Xi_{x,tz},
\Xi_{y,tz}, \Phi_{,tt}, \Theta_{,tt} ) .
\end{align}
\end{widetext}
Here, $f_i$ and $g_i$ ($i=1,2,3,4,5$) are linear combinations of the variables in the bracket.
Substituting Eq. \eqref{elim1} or Eq. \eqref{elim2} into the remaining components of Eq. \eqref{Bumvfieldo} and Eq. \eqref{Bumgfieldo}, we get the set of equations as follows:
\begin{widetext}
\begin{itemize}
\item $x$ and $y$ component of vector field equation
\begin{align}
\label{bbbvx}
0=& -\Sigma_{x,tt} +\Sigma_{x,zz}
-b^t \Xi_{x,tt} 
-\frac{1}{2} \rho b^z  \Xi_{x,tz} +(1-\rho/2)b^t\Xi_{x,zz} \nonumber \\
&+\frac12 b^x[ -2(1+\rho )\Theta_{,tt} +(2+\rho)\Theta_{,zz} 
 +2\rho \Phi_{,zz}] 
 \nonumber \\
&+\frac12 (2-\rho) b^x  (-h_{+,tt} + h_{+,zz})
 +\frac12 (2-\rho) b^y (-h_{\times,tt}+ h_{\times,zz})
\equiv \mathcal{G}_{vx} ,
\\
\label{bbbvy}
0=& -\Sigma_{y,tt} +\Sigma_{y,zz}
-b^t \Xi_{y,tt} 
-\frac{1}{2}\rho b^z \Xi_{y,tz} 
+(1-\rho/2)b^t\Xi_{y,zz}  
\nonumber \\
&+\frac12 b^y [-2(1+\rho )\Theta_{,tt} +(2+\rho)\Theta_{,zz} 
+ 2\rho \Phi_{,zz} ] \nonumber \\
&-\frac12 (2-\rho) b^y (-h_{+,tt} + h_{+,zz})
+\frac12 (2-\rho) b^x (-h_{\times,tt}+ h_{\times,zz})
\equiv \mathcal{G}_{vy} ,
\end{align}
\item $tx$ and $ty$ component of metric field equation
\begin{align}
\label{bbbtx}
0= & -2\xi b^t \Sigma_{x,tt} 
-2 \xi b^z \Sigma_{x,tz}
-2\xi b_t^2 \Xi_{x,tt} 
-\xi b^t b^z (2+\rho) \Xi_{x,tz} 
\nonumber \\
&-[2+ \xi(2-\rho)(-b_t^2 +b_x^2) +2\xi b_z^2] \Xi_{x,zz}
-\xi b^x b^y (2-\rho)\Xi_{y,zz}
\nonumber \\
& +\xi b^x [ -2 (1+\rho)b^t\Theta_{,tt} 
-2  ( 2-\rho ) b^z\Theta_{,tz}
-(2-3\rho) b^t \Theta_{,zz}
+2\rho b^t  \Phi_{,zz} ]
\nonumber \\
&+\xi (2 -\rho)b^t b^x(-h_{+,tt} +h_{+,zz})
+\xi (2 -\rho) b^t b^y(-h_{\times,tt} +h_{\times,zz})
\equiv \mathcal{G}_{tx} ,
\\
\label{bbbty}
0= & -2\xi b^t \Sigma_{y,tt} 
-2\xi b^z\Sigma_{y,tz}
-2\xi b_t^2 \Xi_{y,tt} 
-\xi (2+\rho) b^t b^z \Xi_{y,tz} 
\nonumber \\
&-[2+ \xi(2-\rho)(-b_t^2 +b_y^2) +2\xi b_z^2  ] \Xi_{y,zz}
-\xi (2-\rho) b^x b^y \Xi_{x,zz} 
\nonumber \\
& +\xi b^y [-2(1+\rho) b^t\Theta_{,tt} 
-2 ( 2-\rho )  b^z \Theta_{,tz}
-(2-3\rho) b^t \Theta_{,zz}
+2\rho b^t \Phi_{,zz} ] 
\nonumber \\
& -\xi (2 -\rho)b^tb^y(-h_{+,tt} +h_{+,zz})
+\xi (2 -\rho)b^tb^x(-h_{\times,tt} +h_{\times,zz})
\equiv \mathcal{G}_{ty} ,
\end{align}
\item $xz$ and $yz$ component of metric field equation
\begin{align}
\label{bbbxz}
0 =& \hspace{0.2cm} 2\xi b^t \Sigma_{x,tz} 
+2\xi b^z \Sigma_{x,zz}
+[2+ \xi(2-\rho)(b_x^2+b_z^2)]\Xi_{x,tz}
\nonumber \\
& +\xi(2-\rho) b^t b^z\Xi_{x,zz}
+\xi(2-\rho) b^x b^y \Xi_{y,tz} 
\nonumber \\
& +\xi b^x [ 2 (1-2\rho) b^z \Theta_{,tt}
+2(2-\rho) b^t  \Theta_{,tz}
+(2+\rho ) b^z \Theta_{,zz}
+2\rho  b^z \Phi_{,zz} ]
\nonumber \\
&+\xi (2-\rho)b^x b^z(-h_{+,tt}+h_{+,zz})
+\xi (2-\rho) b^y b^z (-h_{\times,tt}+h_{\times,zz})
\equiv \mathcal{G}_{xz}  ,
\\
\label{bbbyz}
0 =& \hspace{0.2cm}  2\xi b^t \Sigma_{y,tz} 
+2\xi b^z \Sigma_{y,zz}
+[1+ \xi (2-\rho)(b_y^2+b_z^2)]\Xi_{y,tz}
\nonumber \\
& +\xi (2-\rho) b^t b^z\Xi_{y,zz}
+\xi (2-\rho) b^x b^y\Xi_{x,tz} 
\nonumber \\
& +\xi b^y[ 2(1-2\rho)b^z \Theta_{,tt}
+ 2(2-\rho) b^t\Theta_{,tz}
+ (2+\rho) b^z \Theta_{,zz}
+ 2\rho b^z \Phi_{,zz} ]
\nonumber \\
&-\xi (2-\rho) b^y b^z (-h_{+,tt}+h_{+,zz})
 +\xi (2-\rho) b^x b^z (-h_{\times,tt}+h_{\times,zz})
\equiv \mathcal{G}_{yz} ,
\end{align}
\item $xy$ component of the metric field equation
\begin{align}
\label{bbbxy}
0 =& \hspace{0.2cm}   \xi (2-\rho) [ b^y b^z \Xi_{x,tz}
+\xi b^x b^z \Xi_{y,tz}
+\xi b^t b^y \Xi_{x,zz}
+\xi b^t b^x \Xi_{y,zz} ] 
\nonumber \\
&+ 2\xi b^xb^y [ (1-2\rho)\Theta_{,tt}
 +\rho \Theta_{,zz}
 -2(1-\rho)\Phi_{,zz}] 
 \nonumber \\
&+ [-2 +2\xi b_t^2 -\xi(2-\rho)(b_x^2+b_y^2)]h_{\times,tt}
+4\xi b^t b^z h_{\times,tz} 
\nonumber \\
&+[2 +2\xi b_z^2 + \xi (2-\rho)(b_x^2+b_y^2)] h_{\times,zz}
\equiv \mathcal{G}_{xy} ,
\end{align}
\item $xx$ minus $yy$ component of the metric field equation
\begin{align}
\label{bbbxmy}
0 =& \hspace{0.2cm}  \xi (2-\rho) [ b^x b^z \Xi_{x,tz}
- b^y b^z \Xi_{y,tz}
+b^t b^x \Xi_{x,zz}
-b^t b^y \Xi_{y,zz}] 
\nonumber \\
&+ \xi (b_x^2-b_y^2) [(1-2\rho)\Theta_{,tt}
+\rho \Theta_{,zz}
-2(1-\rho)\Phi_{,zz}] 
\nonumber \\
&+[-2+2\xi b_t^2 -\xi (2-\rho) (b_x^2+b_y^2)]h_{+,tt}
+4\xi b^t b^z h_{+,tz} 
\nonumber \\
&+[2+2\xi b_z^2 +\xi (2-\rho) (b_x^2+b_y^2)]h_{+,zz}
\equiv \mathcal{G}_{x-y} ,
\end{align}
\item $xx$ plus $yy$ component of the metric field equation
\begin{align}
\label{bbbxpy}
0=& -\xi (2-\rho)[b^x b^z\Xi_{x,tz}
+b^yb^z\Xi_{y,tz}
+b^tb^x\Xi_{x,zz}
+b^tb^y \Xi_{y,zz} ] 
\nonumber \\
& +\xi (1-2\rho) (b_x^2+b_y^2-2b_z^2 )\Theta_{,tt}
-4 \xi (1-2\rho) b^tb^z \Theta_{,tz}
\nonumber \\
&+[2+\xi (-4b_t^2 +2b_z^2) 
+\xi \rho(4b_t^2 +b_x^2 +b_y^2)] \Theta_{,zz} 
\nonumber \\
&+[ -4 + \xi (4b_t^2-2b_x^2-2b_y^2-4b_z^2)
+2\xi \rho (b_x^2+b_y^2)]\Phi_{,zz} 
\nonumber \\
& +\xi (2-\rho)(b_x^2-b_y^2)
(-h_{+,tt}+h_{+,zz})
+2b^xb^y\xi(2-\rho)(-h_{\times,tt}
+h_{\times,zz})
\equiv \mathcal{G}_{x+y} .
\end{align}
\end{itemize}
The conservation condition \eqref{vprime} becomes
\begin{align}
\label{bbbc}
0 &= \lambda \xi \bigg[(b_x^2-b_y^2)h_+
+2b^x b^y h_\times
+ 2b^x (\Sigma_{x} +b^t \Xi_x)
+2b^y (\Sigma_y +b^t \Xi_y) 
\nonumber \\
& \hspace{0.5cm}
+\left[-2 +\xi (2b_t^2 +b_x^2 +b_y^2 -2b_z^2) \right] \Theta \bigg] 
+\frac{\xi \rho}{4}(2\Phi_{,zz} +2\Theta_{,zz} -3\Theta_{,tt} )\equiv \mathcal{G}_c .
\end{align}
\end{widetext}
Notice that, all the $\Psi$-term and $\Omega$-term vanish. 
If $b^x=b^y=0$, the tensor, vector, and scalar perturbations decouple directly, and we can easily get the results in Sec. \ref{bxby0}.
For $b_x^2+b_y^2\neq0$, in order to separate the perturbations, we take two derivatives with respect to $t$ or $z$ of Eqs. \eqref{bbbvx} to \eqref{bbbc}, i.e.
\begin{align}
\begin{array}{lll}
\partial_t^2 \mathcal{G}_{vx}=0, & 
\partial_t\partial_z \mathcal{G}_{vx}=0,  & 
\partial_z^2 \mathcal{G}_{vx}=0,
\\
\partial_t^2 \mathcal{G}_{vy}=0, &
\partial_t\partial_z \mathcal{G}_{vy}=0, &
\partial_z^2 \mathcal{G}_{vy}=0,  
\\
\partial_t^2 \mathcal{G}_{tx}=0, &
\partial_t\partial_z \mathcal{G}_{tx}=0, &
\partial_z^2 \mathcal{G}_{tx}=0, 
\\
\partial_t^2 \mathcal{G}_{ty}=0, &
\partial_t\partial_z \mathcal{G}_{ty}=0, &
\partial_z^2 \mathcal{G}_{ty}=0,     
\\
\partial_t^2 \mathcal{G}_{xz}=0, &
\partial_t\partial_z \mathcal{G}_{xz}=0, &
\partial_z^2 \mathcal{G}_{xz}=0,     
\\
\partial_t^2 \mathcal{G}_{yz}=0, &
\partial_t\partial_z \mathcal{G}_{yz}=0, & \partial_z^2 \mathcal{G}_{yz}=0,     
\\
\partial_t^2 \mathcal{G}_{xy}=0, &
\partial_t\partial_z \mathcal{G}_{xy}=0, &
\partial_z^2 \mathcal{G}_{xy}=0, 
\\
\partial_t^2 \mathcal{G}_{x-y}=0, &
\partial_t\partial_z \mathcal{G}_{x-y}=0, &
\partial_z^2 \mathcal{G}_{x-y}=0, 
\\
\partial_t^2 \mathcal{G}_{x+y}=0, &
\partial_t\partial_z \mathcal{G}_{x+y}=0, &
\partial_z^2 \mathcal{G}_{x+y}=0, 
\\ 
\partial_t^2 \mathcal{G}_{c}=0, &
\partial_t\partial_z \mathcal{G}_{c}=0,  &
\partial_z^2 \mathcal{G}_{c}=0.   
\end{array}
\end{align}
Under the premise of $b^t\neq b^z$, we repeat the elimination to reduce the variables, and finally get the
four spacetime derivatives equations for $h_+$ or $h_\times$ only. Assuming a wave-like solution, we can get two different sets of dispersion relations, and the results are displayed in Sec.~\ref{bumgeneral}.
We apply similar analysis when $b^t=b^z$, and obtain the results in Sec.~\ref{bt=bz}.

\section{Calculation details in the bumblebee-like SME model}
\label{bsmeappendix}

Different from the original bumblebee model, there is no $\Omega$, $\Psi$, $\Sigma_x$, and $\Sigma_y$ in the linearized metric field equation in the bumblebee-like SME model. 
We first use $tt$, $tz$, $zz$, and $xx+yy$ components of Eq.~\eqref{bSME} to eliminate $\Theta$ and $\Phi$:
\begin{widetext}
\begin{align}
\label{SME-scalar}
\Theta_{,tt} &= -\frac{ \xi (b_x^2-b_y^2) h_{+,tt} 
+2\xi b^x b^y h_{\times,tt} }
{2+\xi (-2b_t^2 +b_x^2+b_y^2+2b_z^2) } ,
\nonumber \\
\Theta_{,tz} &= -\frac{\xi (b_x^2-b_y^2) h_{+,tz}
+2\xi b^x b^y h_{\times,tz} }
{2+\xi (-2b_t^2 +b_x^2+b_y^2+2b_z^2) } ,
\nonumber \\
\Theta_{,zz} &= -\frac{\xi(b_x^2-b_y^2) h_{+,zz} 
+2\xi b^x b^y h_{\times,zz} }
{2+\xi(-2b_t^2 +b_x^2+b_y^2+2b_z^2) } ,
\nonumber \\
\Phi_{,zz} &= \frac{\mathcal{F}_{1}(\Xi_{x,tz}, \Xi_{y,tz},\Xi_{x,zz},\Xi_{y,zz}, h_{+,tt},h_{\times,tt},h_{+,tz},h_{\times,tz},h_{+,zz},h_{\times,zz} ) }
{2[2+ \xi(-2b_t^2 +b_x^2+b_y^2+2b_z^2)]^2 } .
\end{align}
\end{widetext}
Substituting them into Eq. \eqref{bSME}, then we use $tx$, $ty$, $xz$, and $yz$ components to eliminate $\Xi_{x}$ and $\Xi_{y}$:
\begin{align}
\label{SME-vector}
\Xi_{x,tz} &= \frac{ \mathcal{F}_{2}
( h_{+,tt},h_{\times,tt},h_{+,tz},h_{\times,tz} )}
{[1+\xi(-b_t^2+b_z^2)]
[2+\xi (-2b_t^2 +b_x^2+b_y^2+2b_z^2))]} ,
\nonumber \\
\Xi_{y,tz} &=   \frac{ \mathcal{F}_{3}
( h_{+,tt},h_{\times,tt},h_{+,tz},h_{\times,tz} )}
{[1+\xi(-b_t^2+b_z^2)]
[2+\xi(-2b_t^2 +b_x^2+b_y^2+2b_z^2))]} ,
\nonumber \\
\Xi_{x,zz} &=  \frac{ \mathcal{F}_{4}
( h_{+,tz},h_{\times,tz},h_{+,zz},h_{\times,zz} )}
{[1+\xi(-b_t^2+b_z^2)]
[2+\xi(-2b_t^2 +b_x^2+b_y^2+2b_z^2))]} ,
\nonumber \\
\Xi_{y,zz} &=  \frac{ \mathcal{F}_{5}
( h_{+,tz},h_{\times,tz},h_{+,zz},h_{\times,zz} )}
{[1+\xi(-b_t^2+b_z^2)]
[2+\xi(-2b_t^2 +b_x^2+b_y^2+2b_z^2))]} .
\end{align}
Here, $\mathcal{F}_i$ ($i=1,2,3,4,5$) are some linear combinations of the vector and tensor perturbations.
Finally, we substitute Eq.~\eqref{SME-vector} into the $xy$ and $xx-yy$ components of Eq.~\eqref{bSME} and then get the wave equations for $h_+$ and $h_\times$, which are the same as Eq. \eqref{bumso1p} and Eq. \eqref{bumso1c}.
The other perturbations are given by Eq. \eqref{s1th} to Eq. \eqref{s1xiy}.
Since $\big| \xi b^\mu b^\nu \big| \ll 1$, the denominator in the above elimination process is always nonzero.

\appendix


%

\end{document}